\documentclass[12pt]{spieman}  
\usepackage{amsmath,amsfonts,amssymb,mathtools}
\usepackage{graphicx}
\usepackage{setspace}
\usepackage{tocloft}
\usepackage{multirow}
\usepackage{tabularx}
\usepackage{verbatim}

\title{MEMS Deformable Mirror Development for High-Contrast Imaging I: Miniaturized, Flight-Capable Control Electronics} 
\usepackage{array}
\newcolumntype{P}[1]{>{\centering\arraybackslash}p{#1}}

\author[a,*]{Eduardo~Bendek}
\author[a]{Garreth~Ruane}
\author[a]{Camilo~Mejia~Prada}
\author[b]{Christopher B. Mendillo}
\author[a]{A~J~Eldorado~Riggs}
\author[a]{Eugene~Serabyn}

\affil[a]{Jet Propulsion Laboratory, California Institute of Technology, 4800 Oak Grove Dr., Pasadena, CA 91109, USA}
\affil[b]{Lowell Center for Space Science and Technology, University of Massachusetts Lowell, 600 Suffolk Street 3rd Floor, Lowell, MA 01854, USA}

\cftpagenumbersoff{figure}
\cftpagenumbersoff{table}

\graphicspath{{Figures/}}
 
\begin{document} 
  \maketitle 

\begin{abstract}
Deformable mirrors (DMs) are a critical technology to enable coronagraphic direct imaging of exoplanets with current and planned ground- and space-based telescopes as well as future mission concepts, such as the Habitable Exoplanet Observatory (HabEx) and the Large UV/Optical/IR Surveyor (LUVOIR). The latter concepts aim to image exoplanet types ranging from gas giants to Earth analogs. This places several requirements on the DMs such as requires a large actuator count ($\gtrsim$3,000), fine surface height resolution ($\lesssim$10~pm), and radiation hardened driving electronics with low mass and volume. In this paper, we present the design and testing of a flight-capable, miniaturized DM controller. Having achieved contrasts on the order of 5$\times$10$^{-9}$ on a coronagraph testbed in vacuum in the High Contrast Imaging Testbed (HCIT) facility at NASA's Jet Propulsion Laboratory (JPL), we demonstrate that the electronics are capable of meeting the requirements of future coronagraph-equipped space telescopes. We also report on functionality testing on-board the high-altitude balloon experiment ``Planetary Imaging Concept Testbed Using a Recoverable Experiment - Coronagraph (PICTURE-C)," which aims to directly image debris disks and exozodiacal dust around nearby stars. 
The controller is designed for the Boston Micromachines Corporation Kilo-DM and is readily scalable to larger DM formats. 
The three main components of the system (the DM, driving electronics, and mechanical and heat management) are designed to be compact and have low-power consumption to enable its use not only on exoplanet missions, but also in a wide-range of applications that require precision optical systems, such as direct line-of-sight laser communications. The controller is capable of handling 1,024 actuators with 220~V maximum dynamic range, 16-bit resolution, 14-bit accuracy, and 1~kHz operating frequency. The system fits in a 10$\times$10$\times$5~cm$^3$ volume, weighs less than 0.5~kg, and consumes less than 8~W. We have developed a turnkey solution reducing the risk for future missions, lowering their cost by significantly reducing volume, weight, and power consumption of the wavefront control hardware.
\end{abstract}

\keywords{instrumentation, exoplanets, direct detection, coronagraphs, deformable mirrors, control electronics}

{\noindent \footnotesize\textbf{*}Address all correspondance to Eduardo Bendek, Email: \linkable{eduardo.bendek@jpl.nasa.gov}}

\section{Introduction}
\label{sec:intro}  

The direct imaging and spectroscopy of exoplanets in reflected light is one of the next great frontiers in astronomy. Indeed, a primary science goal highlighted in NASA's 30-year roadmap\cite{Kouveliotou2014} is to ``directly image the planets around nearby stars and search their atmospheres for signs of habitability, and perhaps even life.” The Astro2010 decadal survey\cite{Astro2010} reiterated this priority and, as an outcome of the associated recommendations, the Wide Field Infrared Survey Telescope (WFIRST)\cite{Spergel2015} Coronagraph Instrument (CGI)\cite{Noecker2016} will demonstrate key technologies for exoplanet imaging and spectroscopy at visible wavelengths. Furthermore, multiple direct imaging mission concepts with unprecedented scientific potential have been studied, including Exo-C/S\cite{ExoC_finalReport,ExoS_finalReport}, the Habitable Exoplanet Observatory (HabEx)\cite{HabEx_finalReport}, and the Large UV/Optical/IR Surveyor (LUVOIR)\cite{LUVOIR_finalReport}. 

The science case to observe Earth-like planets orbiting Solar-type stars requires image contrasts in visible light on the order of 10$^{-10}$ at sub-arcsecond separations from the host star. A leading approach is to use an internal coronagraph with high-precision wavefront control to suppress unwanted starlight at the position of the planet. A deformable mirror~(DM) is used actively to correct inevitable wavefront errors in the telescope and instrument optics, which diffract starlight throughout the image plane. If left uncorrected, speckle noise due to the star overwhelms the faint planet signal by orders of magnitude. As as result, the DM must provide picometer-level control and stability\cite{TraubOppenheimer2010,Ruane2020}. Despite these challenging requirements, image contrast that is commensurate with directly detecting Earth-like exoplanets has been demonstrated in a laboratory environment\cite{Trauger2007,Seo2019}. Meanwhile, development of DMs that can deliver the desired contrast and stability in a space environment has been identified as a high-priority Coronagraph Technology Gap\cite{ExEPTechPlanAppendix2019} by NASA’s Exoplanet Exploration Program (ExEP). 

In the following, we report on the development of miniaturized, flight-capable DM control electronics (herein, the ``Mini" controller) that meet the requirements of future space mission concepts designed to image Earth-like planets orbiting Solar-type stars. Specifically, we present the design, implementation, and component-level testing. We also describe system-level validation tests on a coronagraph testbed in vacuum within the High Contrast Imaging Testbed (HCIT) facility at NASA's Jet Propulsion Laboratory (JPL) and as part of the high-altitude balloon mission known as Planetary Imaging Concept Testbed Using a Recoverable Experiment - Coronagraph (PICTURE-C)\cite{Cook2015}.

\begin{figure}[t]
    \centering
    \includegraphics[width=8cm]{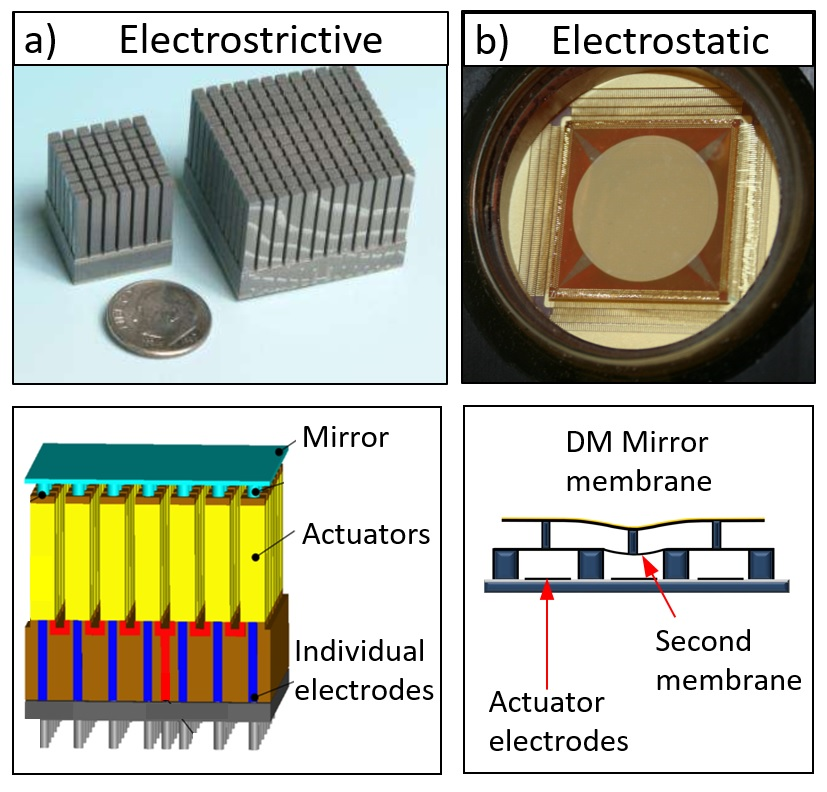}
    \caption{Comparison of electrostrictive and electrostatic DM technologies. (a) Top, real example of electrostrictive actuators arranged in a rectangular matrix. Bottom, electrostrictive DM typical architecture. The mirror is attached to the top of the actuator array which is connected to independent electrodes.  (b) Top, BMC Kilo MEMS DM with electrostatic actuators. Bottom, schematic of the MEMS DM structure composed of a double membrane. Figures adopted from \cite{Cavaco2014} and Boston Micromachines Corporation web page.}
    \label{fig:dm_tech}
\end{figure}

\subsection{Deformable mirror technologies}
There are two main DM actuator technologies currently being considered for space missions (see Fig.~\ref{fig:dm_tech}). The first one is electrostrictive, in which an actuator is mechanically connected to the DM substrate with the mirror membrane\cite{Ealey2004,Wirth2013}. When a voltage is applied to the actuator, it contracts and modifies the mirror surface. The second technology is the electrostatically-forced Micro Electro Mechanical System (MEMS) DM\cite{Bifano2011,Morgan2019}. In this case, the mirror membrane is deformed by a electrostatic force between an electrode where the command voltage is applied and the membrane which is grounded. Thus, the MEMS DMs are contactless, which reduces the likelihood of uncommanded motion caused by environmental factors. 

Manufacturing the driver electronics for either technology is challenging because they require controlling many high voltage channels, potentially at high speed. However, the electrostrictive actuators act as capacitors, and every time an actuator is moved there is a “non-negligible” current. On the other hand, the MEMS DM uses an electrostatic force, dramatically reducing the capacitance of each actuator, so there is a negligible current associated with an actuator motion. The difference between these approaches becomes more important as the number of actuators grows because of the limited ability to miniaturize the electrostrictive technology, whereas the electrostatic MEMS DMs are manufactured using photolithography.

There are several MEMS DM formats available. The controller described in this paper is designed for the Boston Micromachines Corporation (BMC) Kilo-DM, and is compatible with the 32$\times$32 square array and the 952-actuator (34 across) circular array versions. In both cases, the actuators have 300~$\mu$m pitch and maximum deflection on the order of 1.5~$\mu$m at $\sim$200~V. The maximum stroke and gain varies for each device because the distance from the membrane to the active electrode and its thickness varies due to the manufacturing process. 

\begin{table}[t]
\centering
\begin{tabular}{|c|c|c|}
    \hline
     \bf{Attribute} & \bf{Ground-based HCI} & \bf{Space-based HCI}\\
     \hline
    Primary goal & High Strehl & High Contrast\\ 
    Correction type & Large phase errors  & Faint focal plane speckles \\ 
    DM stroke & Large $\sim$1~$\mu$m & Small ($\leq$0.1~$\mu$m) \\ 
    Loop speed & Fast ($\geq$100~Hz)  & Slow ($\geq$1~Hz)\\ 
    DM stroke resolution & Medium ($\sim$500~pm)  & High ($\sim$10~pm)\\ 
    Stability & Low & High\\
    \hline
\end{tabular}
\caption{Comparison of conventional AO and high-contrast imaging requirements.}
\label{table:comparison}
\end{table}

\subsection{Wavefront control needs and its applications}

In astronomy, the majority of DMs are utilized in atmospheric correction, or conventional adaptive optics (AO), where the primary goal is to maximize the Strehl ratio by conjugating large, low-order (tip, tilt, defocus, etc.) and smaller, mid-spatial frequency phase aberrations. This usually means that the DM must be capable of producing high strokes at high speeds. On the other hand, extremely high contrast imaging in space does not necessarily need high speed or large stroke. Table~\ref{table:comparison} outlines the comparison between these two applications. The three most important DM characteristics for High-Contrast Imaging (HCI) applications are: (1)~the number of actuators, which is proportional to correctable field of view; (2)~the surface height resolution that defines how accurate the DM can control the wavefront, and (3)~the stability of the DM surface. Despite these differences in requirements, high-contrast applications typically use DMs developed for conventional AO because of their availability. However, it is important to design a DM controller that is optimal for the operational regime in which the DM will be used.

\subsection{Available control electronics}
The majority of coronagraph testbeds for space applications have DMs with $\gtrsim$1,000 actuators and stroke on the order of $\sim$0.5-1.5~$\mu$m before flattening. The maximum driving voltages are approximately 100-200~V with a resolution of at least 14~bits over the dynamic range. The current commercial controllers that are capable of meeting these requirements are bulky, heavy, and high-power consuming devices, which doesn't present a major issue for laboratory development, but is a potential showstopper for space-based applications. 
The BMC Kilo DM, for example, is currently one of the most widespread wavefront control solutions for laboratory coronagraphs and ground-based AO instruments. The commercial BMC controller for the Kilo DM utilizes a 19” chassis (5.25”$\times$19”$\times$14”) that consumes up to 40~W and weighs $\sim$10~kg. 
A PCI card installed in a desktop computer communicates with the DM controller through a ribbon cable or a fiber. The resulting high-voltage commands are connected to a printed circuit board (PCB) using several flex cables, which are soldered to a PCB that has a zero insertion force (ZIF) socket where the DM is mounted. However, the controller is not vacuum compatible, so a vacuum feedthrough that uses MEG-array connectors must be used to connect the air and vacuum side of the flex cables. 
\begin{figure}[t]
    \centering
    \includegraphics[width=13cm]{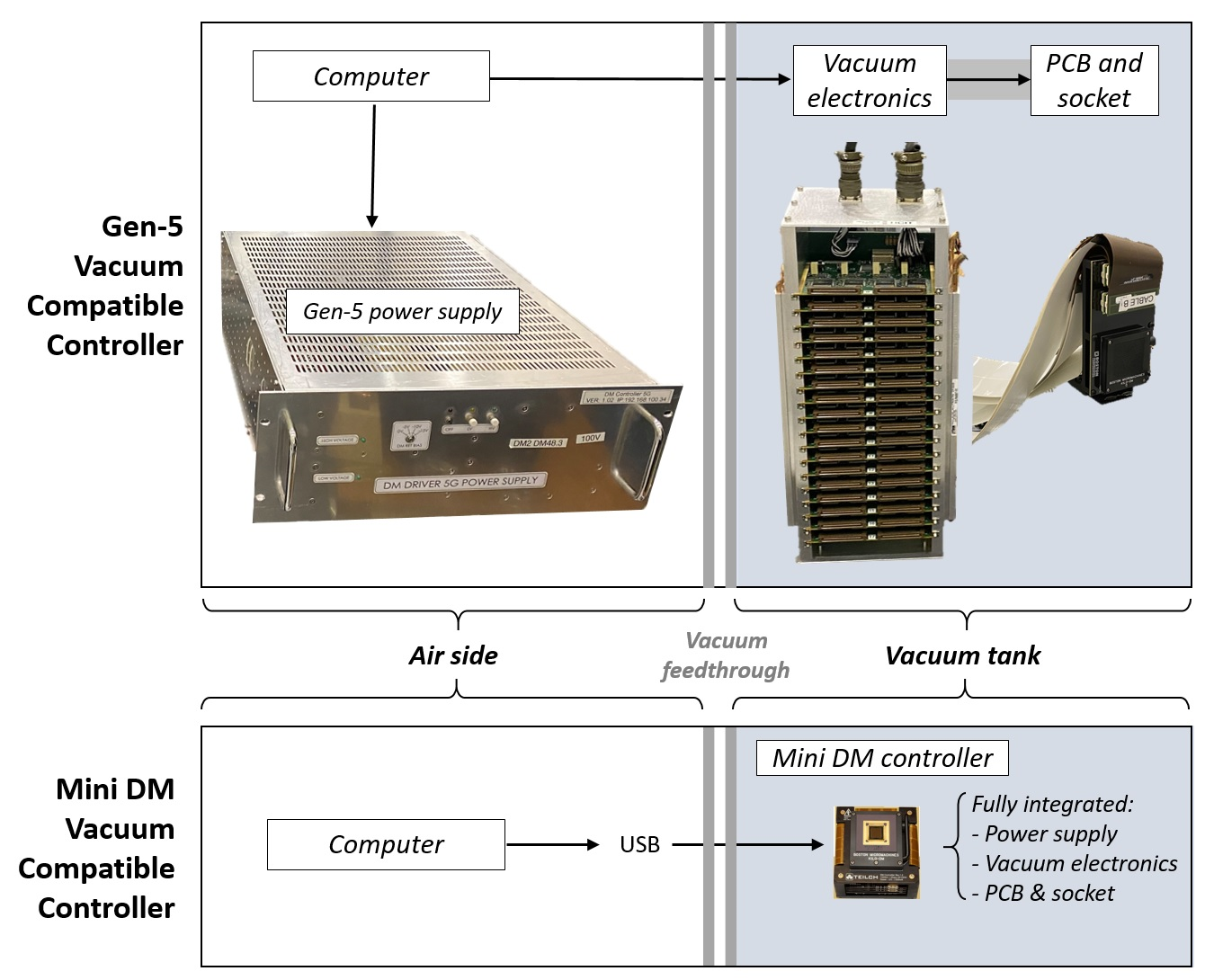}
    \caption{Comparison between the Gen-5 and the Mini DM control electronics. Both are vacuum compatible and offer the same performance. All components are shown in the same scale.}
    \label{fig:Gen-5_mini}
\end{figure}

The JPL HCIT facility has developed custom, vacuum-compatible controllers as part of coronagraph technology development for space missions. A version called Gen-5 is currently in operation at JPL. Table \ref{table:JPL_elec} summarizes the control electronics available on the market or custom made. Figure~\ref{fig:Gen-5_mini} shows a comparison between the Gen-5 control electronics, which consists of three components, and our Mini DM controller. In the case of the Gen-5 electronics, the computer communicates with the power supply box, from which a set of cables enters the vacuum tank using a feedthrough. Inside the tank, the cables connect to the vacuum-side electronics box, and flex cables connect the box to the PCB where the ZIF socket is mounted. In contrast, the Mini DM controller only requires a USB cable and 12V DC power supply. Both units have 16-bit resolution, but the Gen-5 chassis can drive up to two Kilo DMs instead of one.

\begin{table}[]
\centering
\begin{tabular}{|c|c|c|c|}
    \hline
     \bf{Parameter} & \bf{BMC} & \bf{Gen-5} & \bf{Mini} \\
     \hline
    Resolution (bits) & 14 or 16 & 16 & 16\\ 
    Maximum voltage (V) & 220 & 100 & 225\\
    Adjustable dynamic range & No & No & Yes\\
    Vacuum compatible & No & Yes & Yes\\ 
    Components with connectors & 2 & 3 & 1\\ 
    \hline
\end{tabular}
\caption{Comparison of control electronics.}
\label{table:JPL_elec}
\end{table}

\section{Design}\label{sec:design} 
\subsection{Goals and requirements}\label{sec:goals}
The objective of this work is to enable the use of MEMS DMs in space by developing a miniaturized, flight-capable DM controller for exoplanet imaging and other optical applications such as direct line-of-sight laser communications in space. Our solution aims to reduce cost, risk, and development time for those missions. We decided to develop the controller for the BMC Kilo DM because (1)~it is suitable for space applications and high altitude balloon experiments, such as PICTURE-C\cite{Cook2015}; (2)~it is being used in high-contrast imaging laboratories\cite{Mazoyer2019} at NASA Ames\cite{Belikov2016}, Princeton University\cite{Sun2018}, Paris Observatory\cite{Baudoz2018}, Space Telescope Science Institute\cite{Soummer2019}, Caltech\cite{LlopSayson2020}, JPL\cite{MejiaPrada2019}, and others; and (3)~BMC DMs with a similar architecture to the Kilo have been successfully deployed as part of exoplanet imaging instruments on large ground-based telescopes including the Gemini Planet Imager (GPI)\cite{Macintosh2014} and the Subaru Coronagraphic Extreme Adaptive Optics (SCExAO) instrument\cite{Jovanovic2015}. The development of this miniaturized DM controller is based on a prototype developed at NASA Ames Research Center\cite{Bendek2016}.

To meet the needs of future exoplanet imaging space missions, our primary goals are:
\begin{itemize}
    \item Goal 1: To develop a miniaturized controller for the BMC Kilo DM that can fit in a 10x10x10~cm volume with an actuator stroke of 1~$\mu$m.
    \item Goal 2: To develop a scalable DM controller architecture in preparation for large actuator counts (50 to 128 across).
\end{itemize}
Table~\ref{tab:missionreq} summarizes the requirements adopted by WFIRST CGI and future space mission concepts. The contrast and outer working angle (OWA), in particular, are the primary drivers for the component-level requirements for the DM. 
\begin{table}[t]
    \centering
    \begin{tabular}{|c|c|c|c|c|}
        \hline 
        \textbf{Space mission} & \multicolumn{4}{c|}{\textbf{Requirements}} \\ 
        \cline{2-5}
        \textbf{or concept} & Diameter & Contrast & OWA at 550~nm& Actuators across\\
        \hline 
        WFIRST CGI\cite{Noecker2016} & 2.4~m & $\sim$3$\times$10$^{-9}$ & 0.4" (9~$\lambda/D$) & 48\\
        Exo-C\cite{ExoC_finalReport} & 1.3~m & $\sim$10$^{-9}$ & 1.4" (16~$\lambda/D$) & 48\\
        HabEx\cite{HabEx_finalReport} & 4~m & $\sim$10$^{-10}$ & 0.9" (30~$\lambda/D$) & 64 \\
        LUVOIR-B\cite{LUVOIR_finalReport} & 8~m & $\sim$10$^{-10}$ & 0.4" (30~$\lambda/D$) & 64 \\
        LUVOIR-A\cite{LUVOIR_finalReport} & 15~m & $\sim$10$^{-10}$ & 0.5" (60~$\lambda/D$) & 128 \\
        \hline
    \end{tabular}
    \caption{Requirements adopted by WFIRST CGI and the future space mission concepts: Exo-C, HabEx, and LUVOIR.}
    \label{tab:missionreq}
\end{table}
The development of the Mini DM controller aims to fill a gap between the currently-available, bulky electronics and the ambitious needs of future flagship-class exoplanet missions in a time- and cost-effective way. Thus, we derived the controller requirements while considering the Kilo DM parameters, the potential needs of near-term missions, and scalability for future flagship missions. Table \ref{table:requirements} provides a more detailed listing of our assumed requirements for this study. For example, the requirement of having a dark zone out to $>$12~$\lambda/D$ allow us to access the habitable zone and ice-line of F, G, and K stars within 5~pc with $\le$1~m telescope, thus enabling a powerful science case. Temporal response requirements are driven by jitter suppression needs, and to enable ground based AO applications. We specify a command voltage rate change of 40~V/ms, and a refresh rate of 1~kHz that specifies how often new commands are sent. As a result, the Mini DM controller refresh rate is 1~Khz for strokes of 40~V or less with respect to the previous command. Similar rationale has been followed for the other top-level requirements described on Table \ref{table:requirements}. The only requirement that has not be implemented is radiation resilience because of lack of rad-hard ASICS. However, for balloons, low Earth orbit (LEO), and short space missions, shielding could allow the controller to operate in those environments. 

\subsection{Architecture}

\begin{table}[t]
\centering
\begin{tabularx}{\textwidth}{|X|X|X|X|}
    \hline
    \bf{Top-level} & \multirow{2}{4em}{\bf{Rationale}} & \bf{Derived} & \multirow{2}{4em}{\bf{Value}}\\
    \bf{requirement} &  & \bf{requirement} & \\
    \hline
    Create a dark zone at out to $>$12~$\lambda/D$ & Habitable zone of F, G, and K stars within 5~pc with a $\le$1~m telescope & $>$24 actuators across the pupil & Kilo DM with 32 or 34 actuators across\\
    \hline
    \multirow{3}{4cm}{DM stroke and resolution to enable raw contrast of $<$10$^{-9}$ at 20\% bandwidth} & \multirow{3}{4cm}{Imaging Earth-sized planets will require $<$10$^{-9}$ raw contrast} & At least 1~$\mu$m stroke	& 1.2~$\mu$m stroke; 200~V dynamic range \\
    \cline{3-4}
	 & & Actuator resolution of $<$0.1~nm & 70pm (14-bit) precision, 18pm (16-bit) resolution\\
    \cline{3-4}
	 & & DM position stability $<$0.1~$\mu$m & 0.1~$\mu$m or less thermal expansion \\
    \hline
    \multirow{2}{4cm}{Wavefront and jitter control} & \multirow{2}{4cm}{Use one DM for both functions} & Command voltage change rate 40 V/ms & 50 V/ms \\
    \cline{3-4}
    & & Refresh rate of 1kHz & 1kHz \\
    \hline
    Space-qualified & Use on future space telescopes & LEO radiation levels for $>$2 years & TID $>$Not implemented \\
    \hline
    
    \multirow{3}{3cm}{Space-capable} & \multirow{3}{3cm}{Use on Balloons, LEO missions} & Vacuum compatible & yes \\
    \cline{3-4}
     & & Temp. -30C to 50C & -30C to 50C \\
    \cline{3-4}
     & & Shock and vibe & not implemented \\
    \hline
    
    \multirow{3}{3cm}{Cubesat compatible} & \multirow{3}{3cm}{Parts commonality with missions} & Mass $<$1~kg & $\sim$0.5~kg \\
    \cline{3-4}
     & & Volume $<$1U & 10$\times$10$\times$5~cm$^3$ \\
    \cline{3-4}
     & & Power $<$10~W & 8~W~max\\
    \hline
\end{tabularx}
\caption{Deformable mirror controller requirements traceability matrix. }
\label{table:requirements}
\end{table}

\begin{figure}[t]
    \centering
    \includegraphics[width=\linewidth]{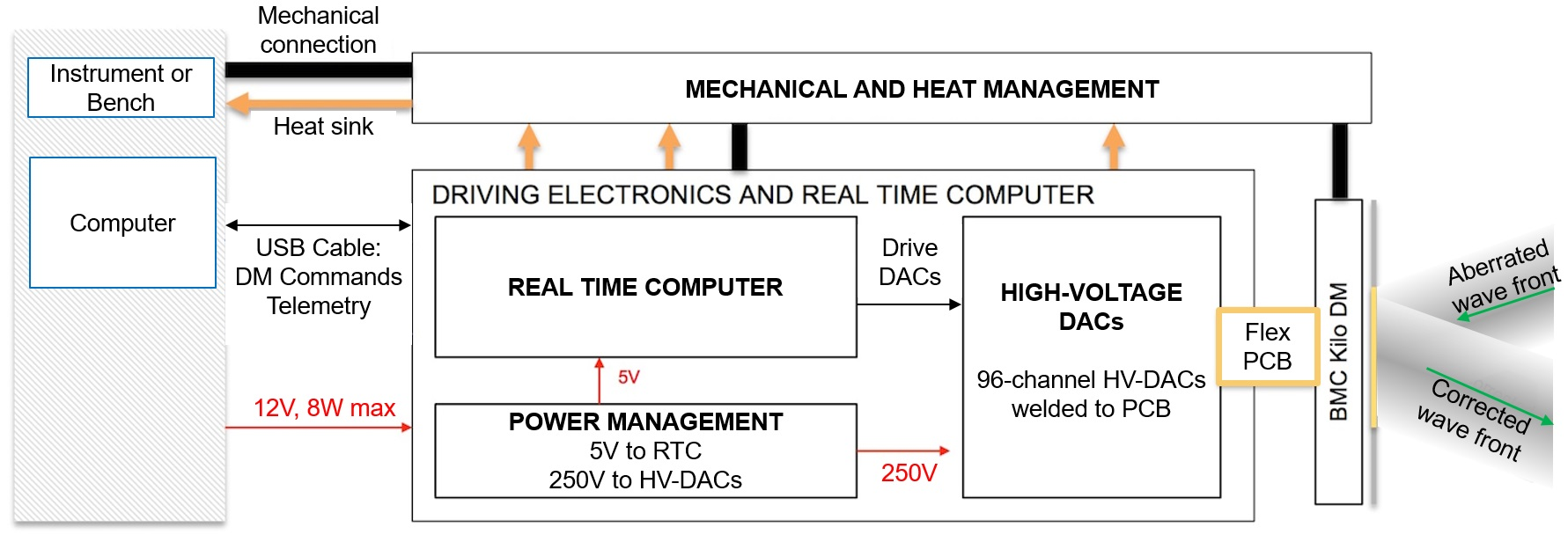}
    \caption{DM Controller functional block diagram.}
    \label{fig:functional_diagram}
\end{figure}

We developed a system to accomplish the goals specified in section \ref{sec:goals}. The Mini DM controller has a volume of 10$\times$10$\times$5~cm$^3$ and weighs less than 0.5~kg. The system has three main functionalities: the DM, which is an off-the-shelf component provided by BMC, the driving electronics (DE) and real time computer (RTC), and the mechanical and heat management (MHM). Figure~\ref{fig:functional_diagram} shows the controller's functional block diagram. The DE provides the communications between the instrument or control computer and the DM using a USB~3 interface, which transmits the new voltage to be applied to each actuator using 32~bit 32$\times$32 (or 34$\times$34) FITS files at 1~kHz requiring a data rate of 4.2~MB/s. 
In addition, there is a telemetry communication channel providing information about the actual voltages applied, system temperature, and power consumption, which uses the same USB cable. The 12~V power supply has a maximum consumption of 8~W. The MHM, which provides mechanical support as well as a heat sink for the DE, connects with the instrument with an array of four \#8-32 threads arranged in a 1-inch square. Also, the MHM holds the DM's ZIF socket and connects it directly to the instrument interface with Invar bars to avoid thermal expansion. The DM receives the command voltages directly from the DE, but its position is also determined by the MHM sub-system.

\subsection{Driving electronics and real time computer}

In our experience at JPL, the electronics and connector reliability have been consistent sources of DM malfunctions. On multiple occasions, actuator anomalies were assumed to be a DM problem, when in reality there was a problem with the driving electronics or connectors. The Mini controller also aims to solve this problem and provide 100\% operational channels. The number of actuators and the number of connections drives the controllable channel yield. Therefore, our design approach removes all connections after the USB input except for the ZIF socket to the DM.

An Application-Specific Integrated Circuit (ASIC) semiconductor chip would be the ideal solution, but there is no ASIC available on the market that can convert 1,024 channels transmitting 16-bit digital TTL signals into 0-220~V analog signals. Also, it would be too expensive to develop a new ASIC for this application. To solve this, we integrate multiple, smaller, commercially-available ASICs. Our design is based on an array of multi-channel, high-voltage sample and hold devices (HV-DAC), such as the 96-channel Dalsa DH9685A. 

The RTC uses a PIC MZ Series microprocessor that manages the communication with the control computer receiving commands from the user interface, and monitors for new DM files when the controller is set to automatically read a file every millisecond. The RTC converts the FITS files containing the control voltage commands into a digital signals to drive the HV-DACs. The RTC retrieves telemetry from the sensors and sends it to the user interface, and runs the Peltier temperature control loop. The controller has high-voltage power supply, which also controlled by the RTC, that provides a maximum of 240~V to the HV-DACs from the 12~V DC externally supplied to the controller. The high voltage power supply can also be set at 200~V, 150~V and 100~V preserving the 16-bit resolution within the dynamic range selected. This setting allows to trade stroke for voltage resolution.

\begin{figure}[t]
    \centering
    \includegraphics[width=\linewidth]{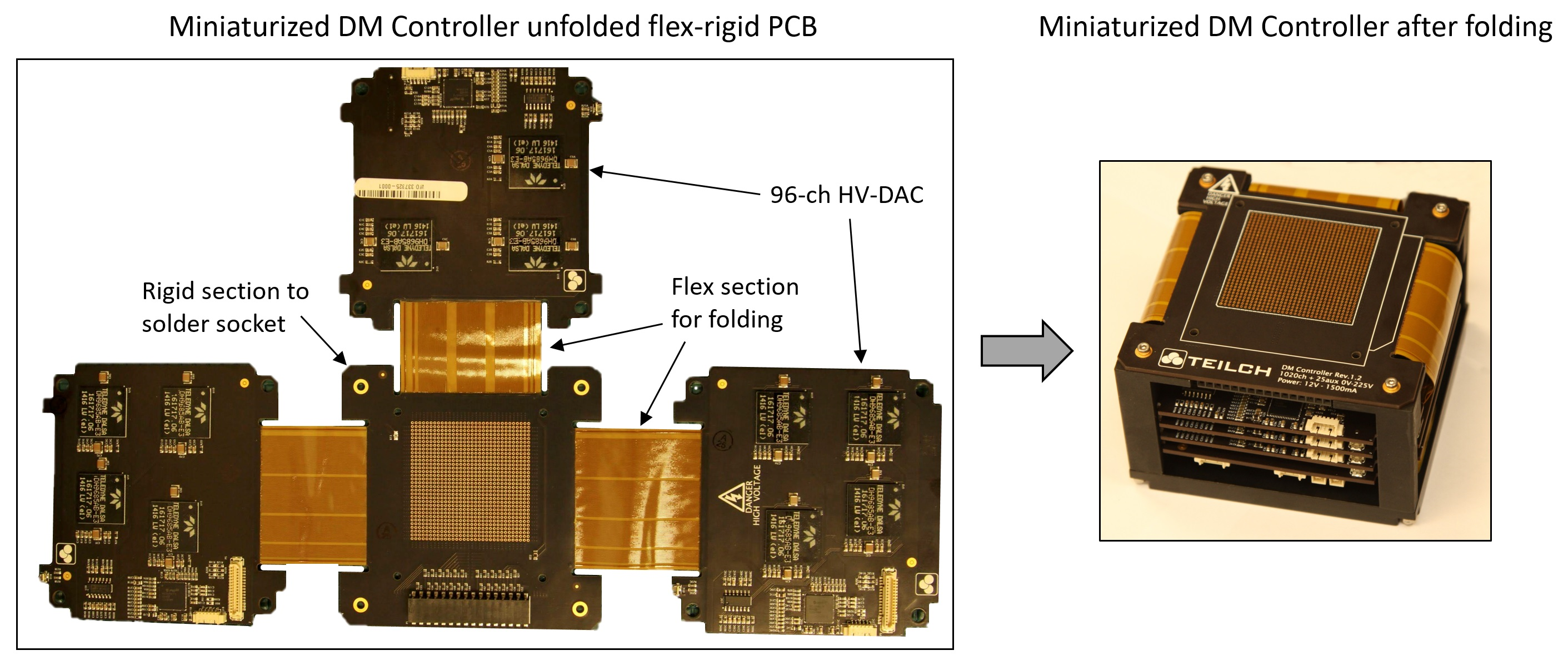}
    \caption{Unfolded and folded images of the flex-rigid PCB.}
    \label{fig:mini_pcb}
\end{figure}

Every electronic component on the controller, including the ZIF socket, is soldered on a single 12-layer flex-rigid PCB to avoid high-density connectors. To fit in a small volume (e.g. a CubeSat format), the flex-rigid PCB folds in four layers that stack on top of each other. The first (central) layer has a rigid section to solder the ZIF socket, which is fed by the other three rigid sections that in total have 11 HV-DACs. Figure~\ref{fig:mini_pcb} shows the flex-rigid PCB. The resulting flex-rigid PCB integrates the HV-DACs, which are high-output impedance analog drivers, with the RTC high-speed digital electronics, and filtered switching power supplies. Special attention was given to the PCB layout and routing to avoid cross-talk and other forms of signal distortion that occur when such diverse electronics are operating on the same PCB. 

\subsection{Mechanical and heat management}

\begin{figure}[t]
    \centering
    \includegraphics[width=\linewidth]{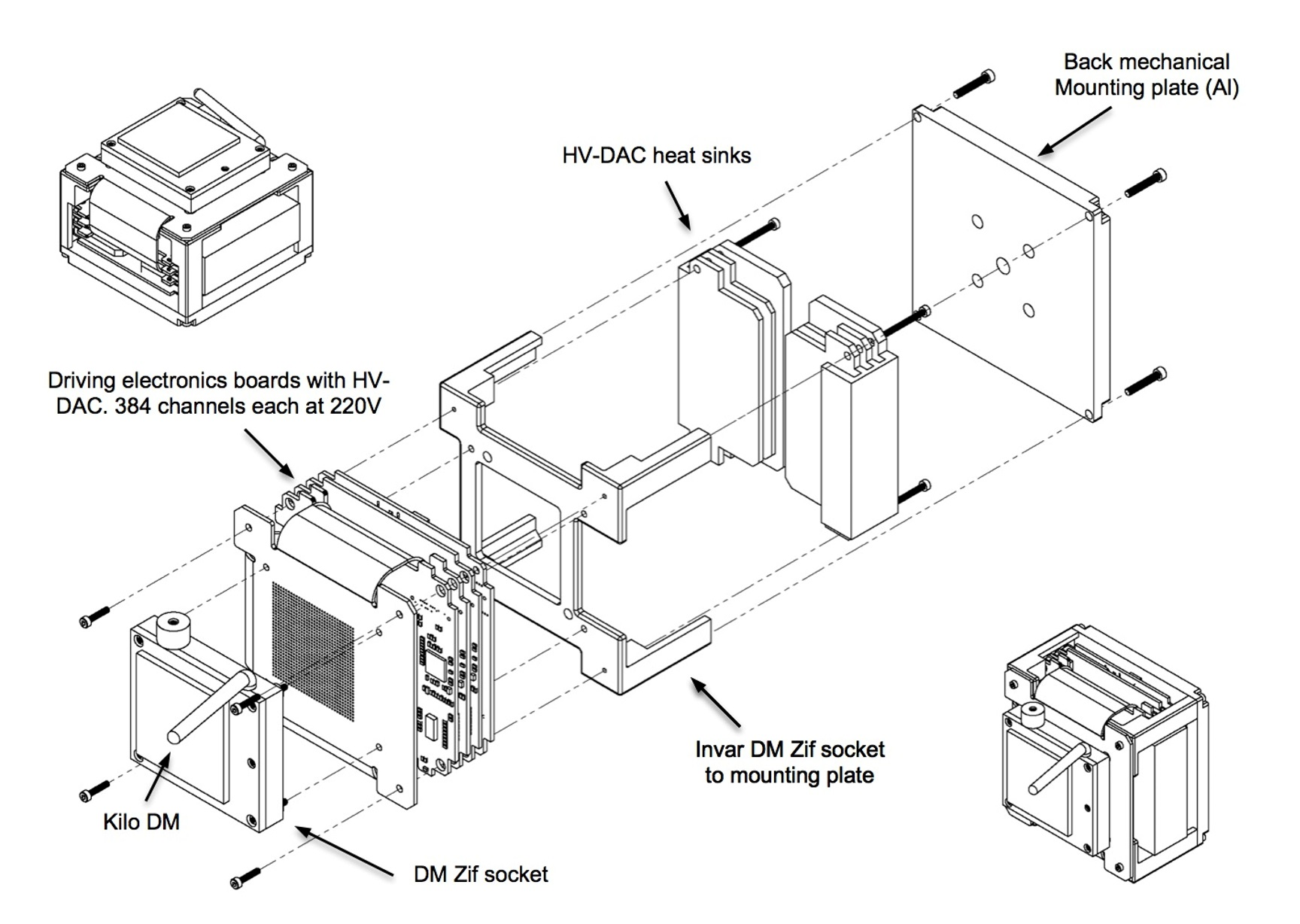}
    \caption{Exploded assembly view of the miniaturized DM controller.}
    \label{fig:mini_mech}
\end{figure}

We also designed the system to provide a stable and rigid mechanical interface with the DM. The Kilo DM is mounted on the ZIF socket custom-made for BMC. The ZIF socket is mechanically attached to an Invar plate, which has 4 legs going around the driving electronics (see Fig.~\ref{fig:mini_mech}). The legs are mounted on an Aluminum standard mounting plate to match CTE using an Aluminum plate that has a standard bolt pattern. The legs are made of Invar to minimize DM tilt induced by differential thermal expansion. The legs are compliant in the lateral direction to absorb the expansion of the connected aluminum plate. There flex PCBs connect the socket PCB with the other PCBs, thus relieving any stress caused by the Invar mounting system. 

The high-voltage power supply and each of the HV-DACs dissipates most of the power consumed by the controller. In the absence of convection, heat sinks that can remove the heat by conduction are necessary. Metallic heat conductors on top of every HV-DAC and the high-voltage power supply are attached to a central heat pipe connected to the instrument to mitigate overheating.

\subsection{Interface and control software}

The controller requires only two connections: data and power. The data is sent to the DM controller through the USB~3 cable. The DM commands are transmitted as 32-bit FITS images of 32$\times$32 (or 34$\times$34) pixels. The value of each pixel represents the voltage to be applied to the respective actuator. The controller can operate at up to 1~kHz, which requires a data rate of 4.2~MB/s. The USB~3 interface also transmits real-time telemetry from the controller at 20~kB/s. Power is supplied using a standard 12~V connector. A PIC microprocessor, running in real-time, reads the images and sends the digital values to the HV-DAC for analog conversion and high-voltage amplification. The microprocessor also controls the high-voltage power supply and reads telemetry. 

On the user side, Microsoft Windows software controls the system through a COM port. Once the system is connected, the user selects the desired FITS file and, if desired, selects the continuous update check box to enable the system to continuously read the file at 1~kHz frequency. The control software overwrites the file to send a new DM command. A flag to make sure the file write was completed avoids conflicts in the reading process.

\subsection{Other design considerations for a space environment}
In order to call a piece of hardware “space qualified,” it must be shown that the hardware can survive and operate in a space environment. While specific missions may have unique environmental concerns, there are established guidelines for general mission planning purposes. The document that NASA utilizes for this purpose is the GSFC-STD-7000A “General Environmental Verification Standard (GEVS)”. Since this project had limited funding ($<$\$150K, including two flight capable units), our approach was to develop a system that is flight-capable, which we define as a unit that can work in vacuum tanks, balloons, and some LEO missions. The unit will not adhere to the stricter GEVS standard.

\subsubsection{The anticipated environment}

The controller design is compatible with space-like conditions including vacuum chambers, balloon missions, and CubeSats. For balloon missions flying at 35~km, the U.S. Standard Atmosphere properties specifies an atmospheric pressure of 7~mbar. The ambient temperature is approximately -30$^\circ$C. These two parameters, atmospheric pressure and temperature, define our operational environment. During the ascent, the temperature will decrease to a minimum of -56$^\circ$C between 11~km and 20~km. Thus, the critical survival conditions including safety margin are -60$^\circ$C and 5~mbar pressure, and the operational conditions are between 0 mbar and 1000~mbar and between -30$^\circ$C and 20$^\circ$C. In the case of CubeSats, the pressure is practically zero. The temperature is maintained at ambient (20$^\circ$C) for most LEO CubeSat missions.

\subsubsection{Radiation}
Critical electronics components such as the 96-channel, 16-bit HV-DACs are not available in rad-hard versions, and initiating a rad-hard development would be a multi-million dollar effort beyond the scope of this work. As an alternative, we believe the system can be hardened against measurable Total Integrated Dose (TID) with shielding such as high molecular weight phenolic novolac doped with a tungsten powder and aluminum foil surface passivation layers to ground dielectric charging that may occur. However, the shield design should be customized for each mission as the orbit, spacecraft, and mission duration will change the optimal design.

\subsubsection{Vacuum and high voltage}
MEMS devices can be damaged by corona discharge or electrical breakdown. The voltage that triggers the discharge is a function of the pressure and gap length; this equation is known as Paschen’s law. According to Paschen's equations, the lowest voltage that will trigger a corona discharge is 327~V at pressure distance of 7.5~mbar$\times$mm in air. For a pressure of 5.5~mbar, the most likely distance breakdown distance is 1.37~mm, which is a typical distance between components or tracks of the electronics. Despite the fact that the controller's 250~V power supply will not reach the breakdown voltage, it is close enough that other effects such as contamination could trigger the breakdown at a lower voltage, thereby posing a risk for the system operating in balloon. In contrast, operation in high-vacuum, in a tank or in LEO orbit, is safer from the corona discharge risk perspective.

Placing the system in vacuum has mechanical challenges as well. Air pockets that are not evacuated can cause stresses when the system is under vacuum. Large integrated circuits, such as the HV-DACs, have the highest risk because the forces can be significant due to their large area. 
\begin{figure}[t]
    \centering
    \includegraphics[width=12cm]{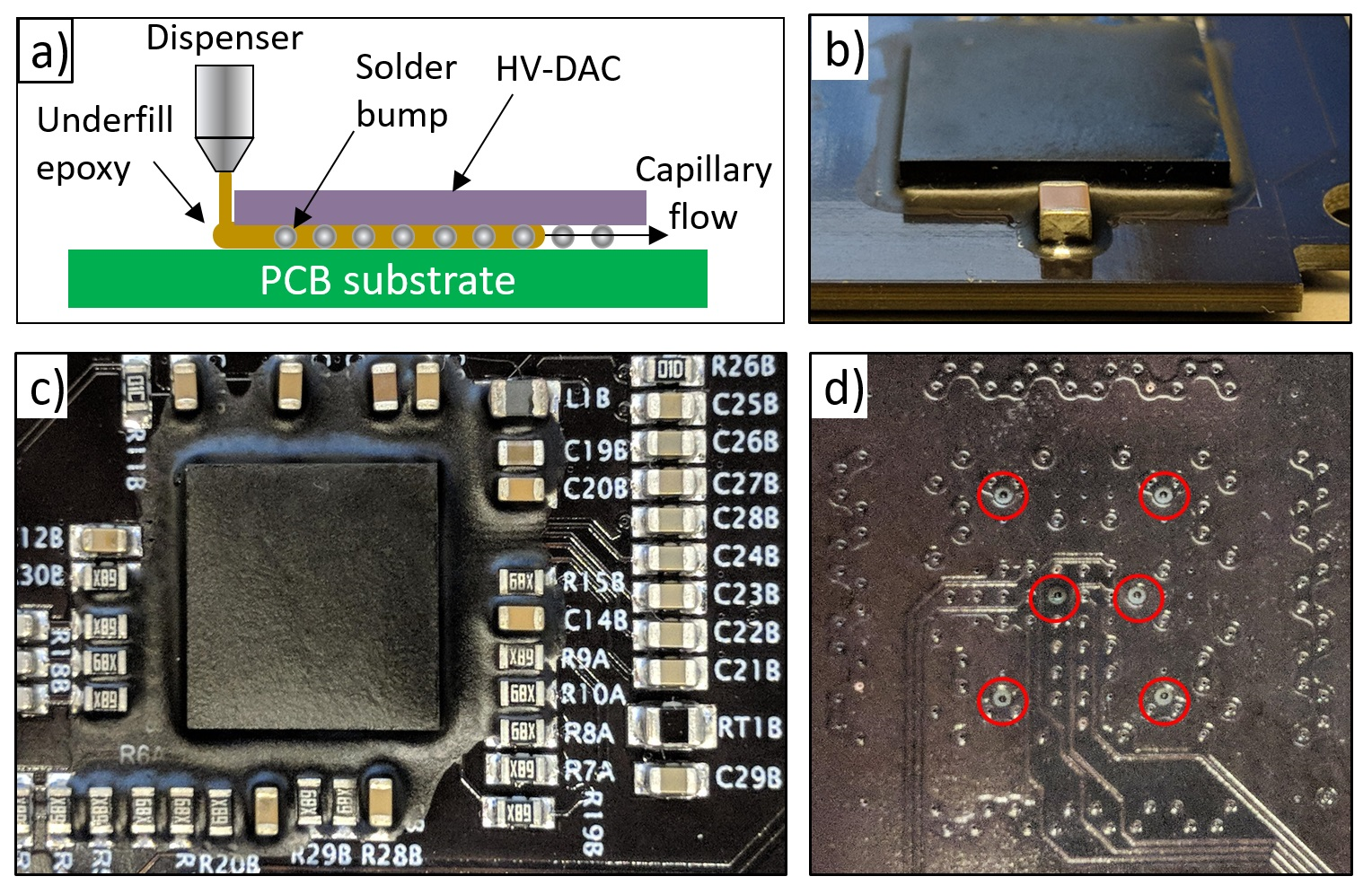}
    \caption{Underfill and coating of the electronics. (a) The underfill application process. (b) An image of a HV-DAC after underfill but before coating. (c) Same as (b), but after acrylic coating. (d) Image of witness holes to verify underfill quality.}
    \label{fig:coatings}
\end{figure}

\subsubsection{Coating of electronic circuits}
To prevent high-voltage breakdown and stress caused by air pockets, two surface coatings were applied to the PCB assemblies for vacuum operation: underfill and conformal coating. The underfill (see Fig.~\ref{fig:coatings}a,b) prevents the formation of air pockets, and increases the reliability by improving mechanical resistance of solder joints to repeated thermal cycles. It also insulates conductive areas under BGA chips to reduce the risk of sparks in high-voltage areas. The underfill material, Loctite Eccobond UF 3810, has a coefficient of thermal expansion (CTE) of 55~ppm/$^\circ$C at temperatures lower than the glass transition at 102$^\circ$C. This CTE is well matched with that of ball solder joints for operation under 102$^\circ$C (PCB and BGA temperature is limited to 85$^\circ$C during normal operation). Capillary holes were added to the PCB to verify the coverage of underfill under BGA chips (5 holes under each BGA). A conformal coating (see Fig.~\ref{fig:coatings}c) was applied after the underfill to provide electrical insulation to exposed high-voltage areas, including socket solder joints. We used MG Chemicals \#419C Acrylic due to its high dielectric strength (450~V/mil, 57~kV/mm). The coating was cured in an oven to minimize outgassing during normal operation. Figure~\ref{fig:coatings}d shows all of the holes under a BGA chip completely filled with underfill material. 

The manufacturing process starts with the PCB delivery, solder paste application, and loading of passive and active electronic components. Then, the solder is reflowed in the oven and the components are hand-loaded and soldered. Next, the DM socket pins are reflowed with soft hot air reflow (300$^\circ$C). Next, the humidity is removed by keeping the boards in oven at 70$^\circ$C for 60~min. Immediately after, the underfill is applied with a syringe and the curing process starts (12~min at 130$^\circ$C). Finally, the conformal coating is applied and it is cured for 90~min at 75$^\circ$C.

\begin{figure}[t]
    \centering
    \includegraphics[width=12cm]{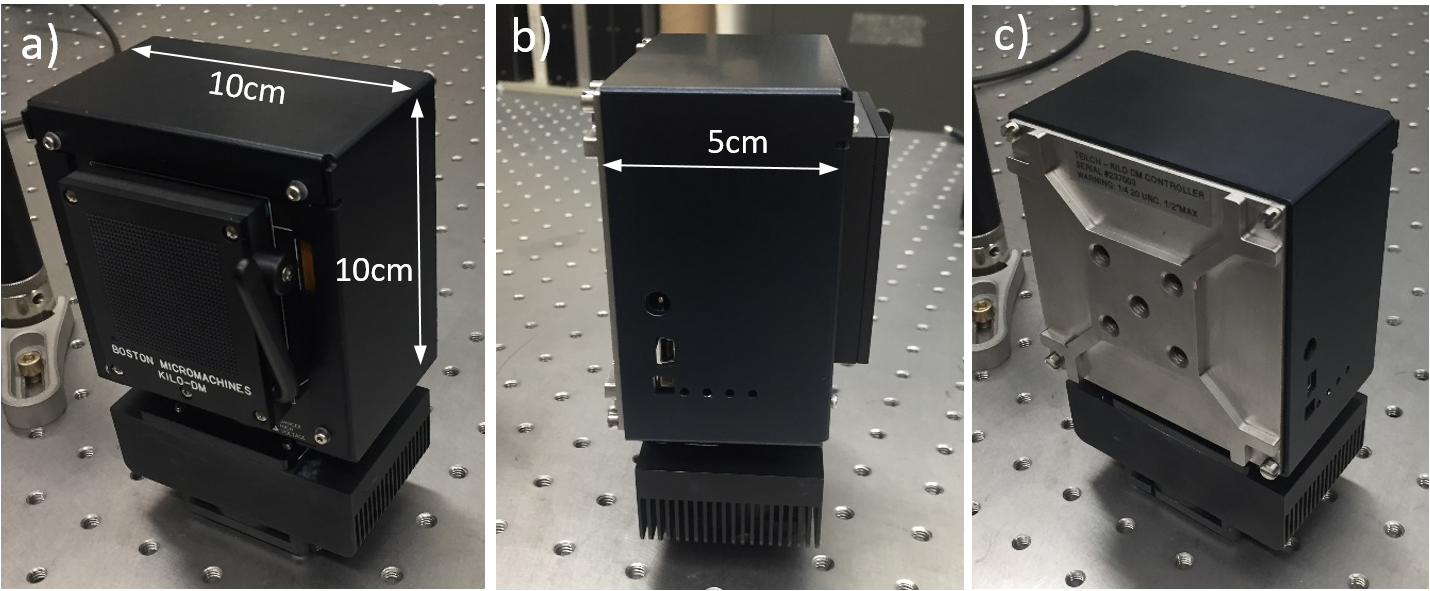}
    \caption{(a)~Front, (b)~side, and (c)~back view of the miniaturized DM controller.}
    \label{fig:controllerpic}
\end{figure}

\subsection{Final product}

Initially we manufactured two identical devices for the PICTURE-C instrument, a flight unit and a spare. The device is shown in Fig.~\ref{fig:controllerpic}. An aluminum heat exchanger with a Peltier and a fan are included for testing in air. For use with coronagraphs in air, we replace the fan with a cold plate to avoid vibrations and convection-induced turbulence. For vacuum operations, a cold plate or copper braid is necessary to avoid overheating 
Each device has 1056 (96$\times$11) available channels, 960 active channels, a maximum dynamic range of 180~V (up to 225~V adjusting HV power supply) with a resolution of 2.7~mV (16~bit), and an accuracy of 10.9~mV (14~bit). The high voltage power supply can also be set at 150~V and 100~V preserving the 16-bit resolution. The maximum rate for DM commands was set at 1~kHz, but can be readily upgraded up to 3~kHz. Finally, the RTC maintains a refresh rate to the HV-DAC of 3~kHz while holding a command. This rate is set to ensure that the control voltage drift is less than $<$1~bit.

\section{Qualification and testing}\label{sec:testing} 

We carried out a series of tests to validate the performance of the DM controller. The first step was to perform electrical functionality tests to verify that the voltages commanded were same as measured on the ZIF socket pins. Then, thermal and vacuum testing was done for operational and survival regimes. After successfully completing those tests, the device performance was characterized using a coronagraph testbed in a vacuum chamber at JPL. Finally, the unit was handed over to the PICTURE-C team for flight testing. 

\begin{figure}[t]
    \centering
    \includegraphics[width=\linewidth]{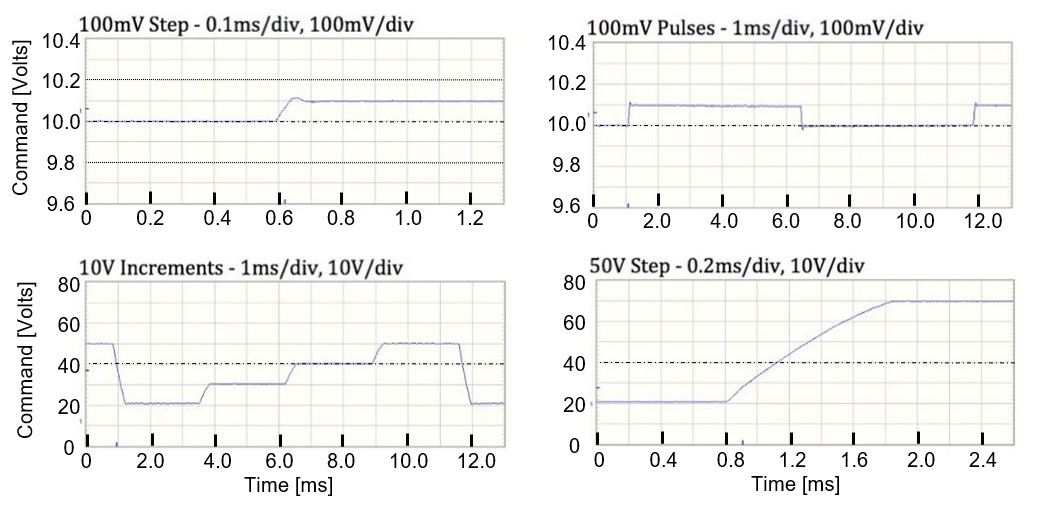}
    \caption{Voltage command tests. (\textit{Top~row})~Command changes of 100~mV take 0.1~ms to stabilize. (\textit{Bottom~row})~The stabilization time is $\le$1.2~ms for commands $\le$50~V. }
    \label{fig:electest}
\end{figure}

\subsection{Electrical functionality}
The MEMS DM's operation is based on the electrostatic force caused by a potential difference between each actuator electrode and the DM membrane, so there is no current circulation through each actuator. Leakage currents are estimated to be less than 1~nA. We validated the commands sent by the DM controller by making voltage measurements directly on the ZIF pins. To do this, the voltmeter or oscilloscope needs $>$1e9~$\Omega$ impedance to avoid loading the HV-DAC and distorting the commands. We connected the oscilloscope to a randomly selected pin and to a ground pin located on each corner to obtain the voltage measurements. Figure~\ref{fig:electest} shows the voltage response versus time for incremental voltage commands of 100~mV, 10~V and 50~V, which take 0.03~ms, 0.3~ms and 1~ms, respectively, with less than 10mV overshoot, and less than 0.1~ms settling time. While the Mini DM controller response time is slower than the BMC DM COTS controller, it is adequate for the intended space-based HCI applications. The DM Controller meets the design requirements in terms of dynamic range of 180~V and 16-bit resolution, which it is equivalent to 2.7~mV for the least significant bit. The noise level is within 10~mV~RMS, meeting the 14-bit accuracy requirement. 
We measured power consumption and voltages after operating the power supply for 1~hr at 200~V in air. In this condition, the power consumption was 6.6~W. Most of the power is dissipated at each 96-channel HV-DAC. Of the 11 of them, their temperatures ranged from 49.8$^\circ$C for the coolest (HV3) to 60.5$^\circ$C for the hottest (HV11). These tests were performed without the heat sink to confirm that the HV-DACs are not at risk of overheating.

\subsection{Survival vacuum and thermal} 

We installed the controller inside a vacuum tank using a feed-through to pass the 12~V DC power supply and the USB data connection. Four 1/2" thick, 8" long copper braids were used to connect the controller heat sink with a large aluminum block that serves as a heat sink. First, we tested the controller operation in vacuum and ambient temperature. We set the power supply to 150~V, and 120~V were applied to all actuators. We evacuated the tank and stabilized the pressure at 6.6~mbar, which is equivalent to 120,000~ft. The telemetry showed stable temperatures and power consumption over 24~hrs.

Afterwards, we tested the controller in vacuum and cold temperatures. With the tank at 6.6~mbar pressure and the controller off, we cooled the tank shroud to -40$^\circ$C and let the temperature stabilize for 24~hrs to an equilibrium temperature of -28$^\circ$C. Then, we turned on the controller with no voltage applied. This configuration consumed 2.6~W and warmed the controller to -20$^\circ$C. The shroud was stabilized at -45$^\circ$C to simulate ambient radiation loads. Next, the power supply was set to 200~V, increasing the power consumption to 4.8~W. We allowed thermalization for 10~min. Afterwards we applied 120~V to all the actuators, increasing consumption to 8.0~W. We kept the system running in this configuration for 30~min. Then, we set voltages back to 0~V, stopped and unplugged controller, and set the cryostat overnight for survival testing.

After the tests the DM was removed and the electrical functionality was verified. Also, an interferometer and a Kilo DM were used to ensure that all actuators were working after the environmental tests. Thus, the DM controller passed the thermal, vacuum, and electrical functionality tests.

\section{System performance characterization at JPL}

In order to test and characterize the performance the DM and electronics as a system, we installed a DM on the ZIF socket, measured the response of each actuator, and then integrated the DM and controller into a coronagraph testbed in a vacuum chamber. Since the assembly needed to efficiently dissipate heat in a vacuum environment, we removed the fan and heat sink assembly from the bottom of the controller (see Fig.~\ref{fig:controllerpic}) and fixed the Peltier cooler directly to an aluminum, liquid-cooled coldplate (MaxQ 001-MXQ-01) with a thin sheet Indium foil in between to enhance the thermal conductivity of the interface. In ambient atmospheric conditions, the controller does not overheat during typical usage with no water flowing through the coldplate. However, in vacuum, flowing water is necessary to dissipate heat efficiently due to the lack of convection. 

\begin{figure}[t]
    \centering
    \includegraphics[height=0.3\linewidth]{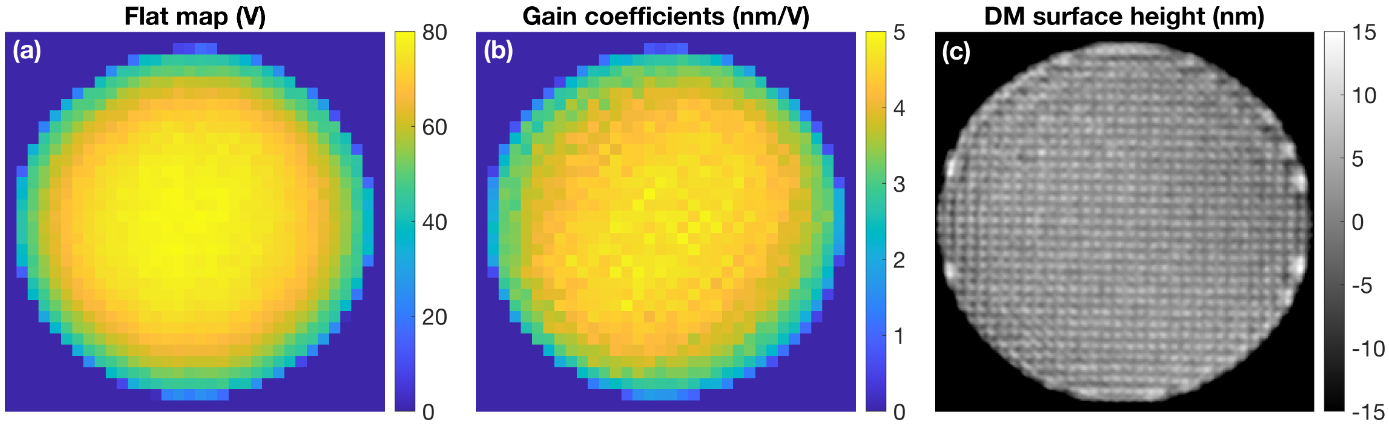}
    \caption{(a)~The voltage map applied to flatten the DM surface. The peak-to-valley is approximately 80~V. (b)~The DM actuator gain coefficients about the flat settings in nm/V. The mean gain is 3.3~nm/V. (c)~The DM surface in nm with the flat map applied measured with a Fizeau interferometer. }
    \label{fig:flat_and_gain}
\end{figure}
\begin{figure}[t]
    \centering
    \includegraphics[height=0.3\linewidth]{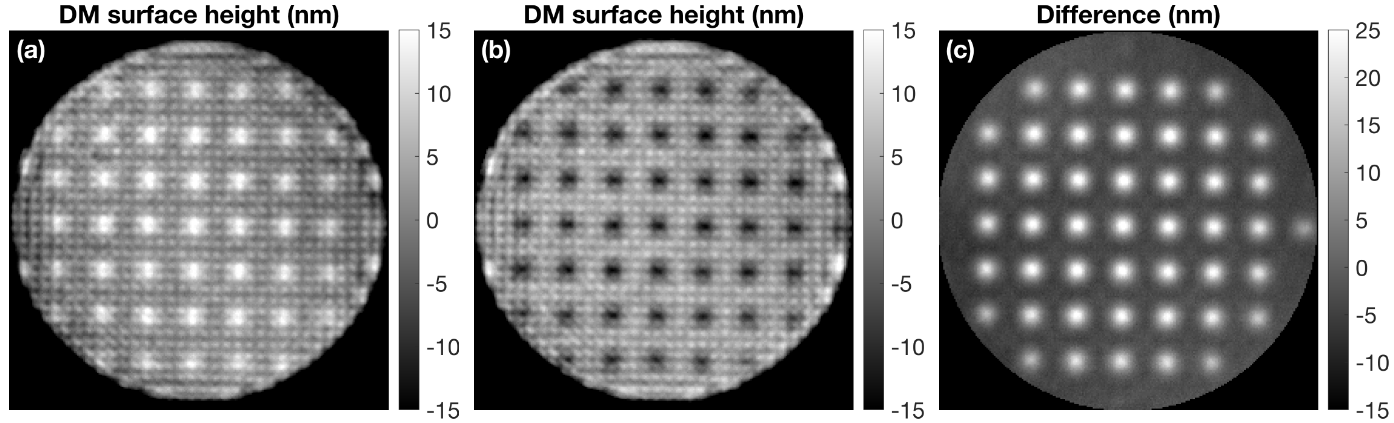}
    \caption{Example DM surface measurements with a grid of poked actuators to determine the gain coefficient of each actuator. (a)-(b)~The DM surface with 1 out of every 16 actuators poked by (a)~-5~V and (b)~+5~V. (c) The difference between (a) and (b) showing the response of the isolated actuators to 10~V difference, from which we estimate the gain coefficients. This measurement is repeated for 16 pairs of measurements to determine the gain of all of the actuators. }
    \label{fig:gain_measurement}
\end{figure}

Prior to installing the DM into the vacuum testbed, we measured the response of each actuator using a Fizeau interferometer (Zygo Verifire). We enclosed the DM inside of a plastic housing with a continuous flow of dry air to maintain a relative humidity $<$30\%, which avoids corrosion on the MEMS DM. The beam from the interferometer passed through a hole in the front of the enclosure. Our typical DM validation and calibration tests consist of exercising each actuator to ensure it responds as expected, flattening the DM, and measuring the gain coefficient (units of nm/V) for each actuator. Since the surface deflection is quadratic with voltage, we measure the gains with the DM at its flat state to best approximate the DM commands in typical scenarios.  Figure~\ref{fig:flat_and_gain} shows the voltages applied to flatten the DM (Fig.~\ref{fig:flat_and_gain}a), the measured gain coefficients (Fig.~\ref{fig:flat_and_gain}b), and the surface of the flattened DM (Fig.~\ref{fig:flat_and_gain}c). The typical DM surface error when the DM is in its flat state is $<$10~nm~RMS over the spatial frequencies resolved by our interferometer. The voltage map that flattens the DM surface (called the flat map) ranges from 0~V near the edges to 80~V near the center. The gain measurement procedure is to poke 1 out of every 16 actuators within a 4$\times$4 square by $\pm$5~V (see Fig.~\ref{fig:gain_measurement}) and then fit the difference of those two measurements to a model of the DM surface. In this way, the gain coefficient for every actuator is determined via 16 pairs of surface measurements. A typical actuator on a BMC Kilo DM has a gain of approximately 4~nm/V in the DM's flat state.

\begin{figure}[t]
    \centering
    \includegraphics[height=0.45\linewidth]{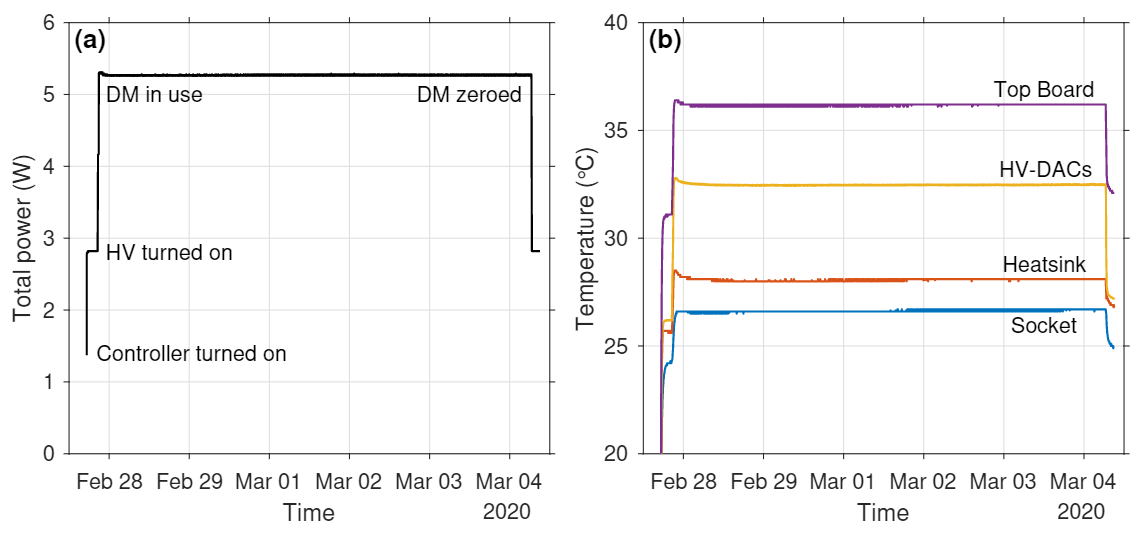}
    \caption{Telemetry from the Mini DM controller during wavefront control experiments in a coronagraph testbed over a 5~day period. (a)~The total power consumption over time with key events labeled. (b)~Temperatures at representative locations on the controller over the same time period. The curve labeled HV-DACs is the mean of the high voltage DACs.}
    \label{fig:PowerTemp}
\end{figure}

After interferometric characterization, we moved the full assembly, including the DM, controller, and cold plate onto a coronagraph testbed. Water hoses were attached to the coldplate and maintained a continuous flow of water at approximately 17$^\circ$C throughout the experiment. The vacuum chamber was kept at pressures of $<$1~mTorr. Using the controller telemetry described above, we monitored the total power and current, the temperature at 19 internal sensors, and the voltage at 18 test points during all of our experiments. Figure~\ref{fig:PowerTemp} shows a subset of the telemetry during wavefront control experiments taking place over a 5-day period. The mean power was 5.27$\pm$0.01~W, where the error quantity is the standard deviation of the measurements over the same time period. Since typical wavefront control results in a consistent level of power consumption, the temperatures stabilize well. The hottest point on the controller is the top board, which is located furthest from the coldplate, and had a temperature of 36.17$\pm$0.06$^\circ$C. The socket was the coolest location, with a temperature of 26.62$\pm$0.05$^\circ$C. And, the heatsink had a temperature of 28.06$\pm$0.06$^\circ$C, which is the closest location to the coldplate and thus the rest of the testbed. The current was also stable at 436$\pm$1~mA. All of the standard deviations reported above were found to be smaller than the resolution of the sensors and therefore are treated as upper limits on the stability.

\begin{figure}[t]
    \centering
    \includegraphics[width=\linewidth]{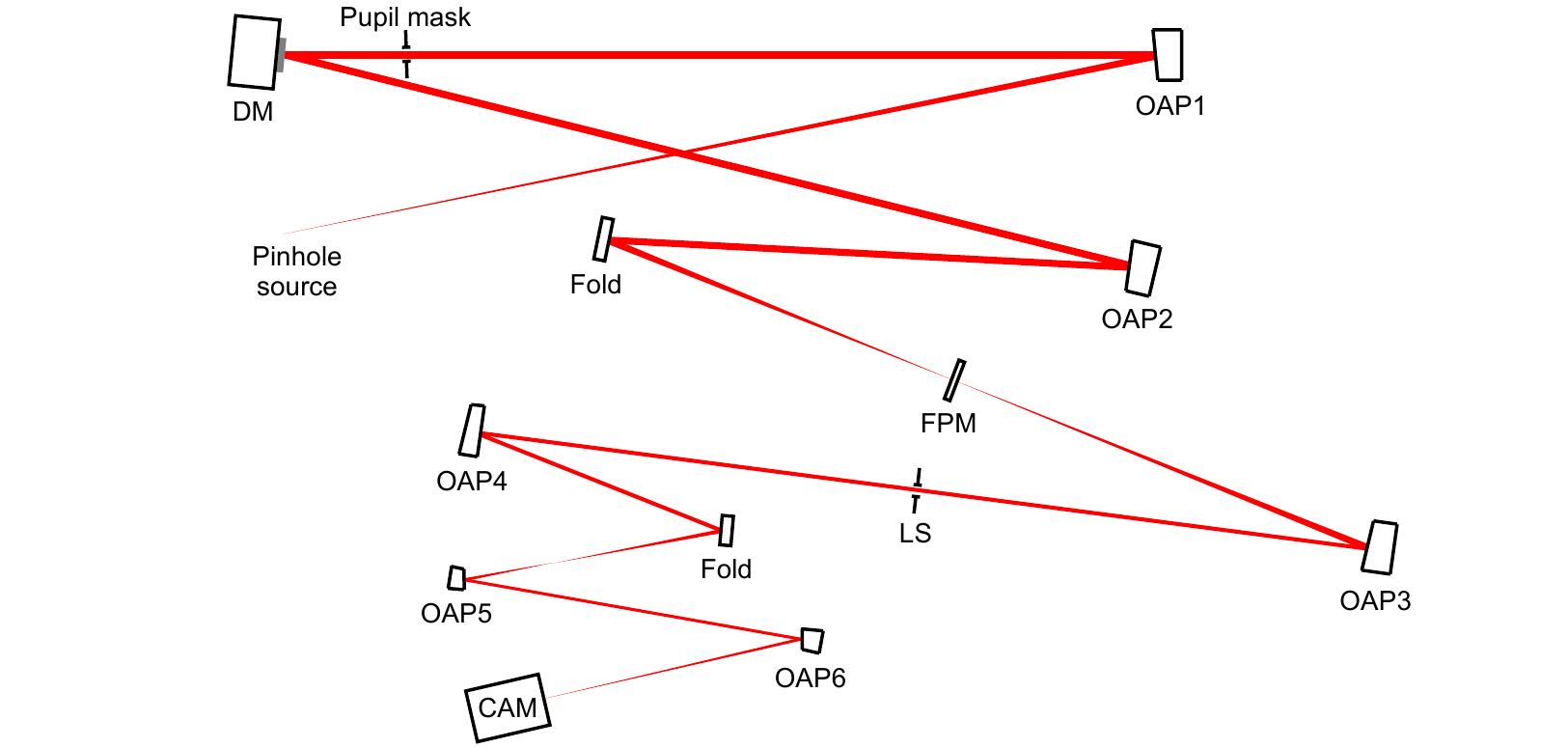}
    \caption{Schematic of the optical system. DM: Deformable mirror. OAP: Off-axis parabola. FPM: Focal plane mask. LS: Lyot stop. CAM: Camera. The relative size of the DM electronics and the path lengths are to scale. }
    \label{fig:optical_schematic}
\end{figure}

The remainder of the testbed was originally designed to demonstrate the performance of vector vortex coronagraphs\cite{Serabyn2019,Ruane2018_JATIS}. Figure~\ref{fig:optical_schematic} shows a schematic of the optical system, where the DM was illuminated by a circular beam created by a supercontinuum laser source to simulate the star and a pupil mask approximately 60~mm in fron\textcolor{green}{t} of the DM. The focal plane mask (FPM) diffracted light for the pseudo-star outside of the Lyot stop (LS). Light that leaks through the LS appears as speckles in the final image, which are suppressed using focal plane wavefront sensing\cite{Giveon2011} and electric field conjugation\cite{Giveon2009,Groff2015} implemented in the open-source FALCO software package\cite{Riggs2018}. 

\begin{figure}[t]
    \centering
    \includegraphics[width=0.5\linewidth]{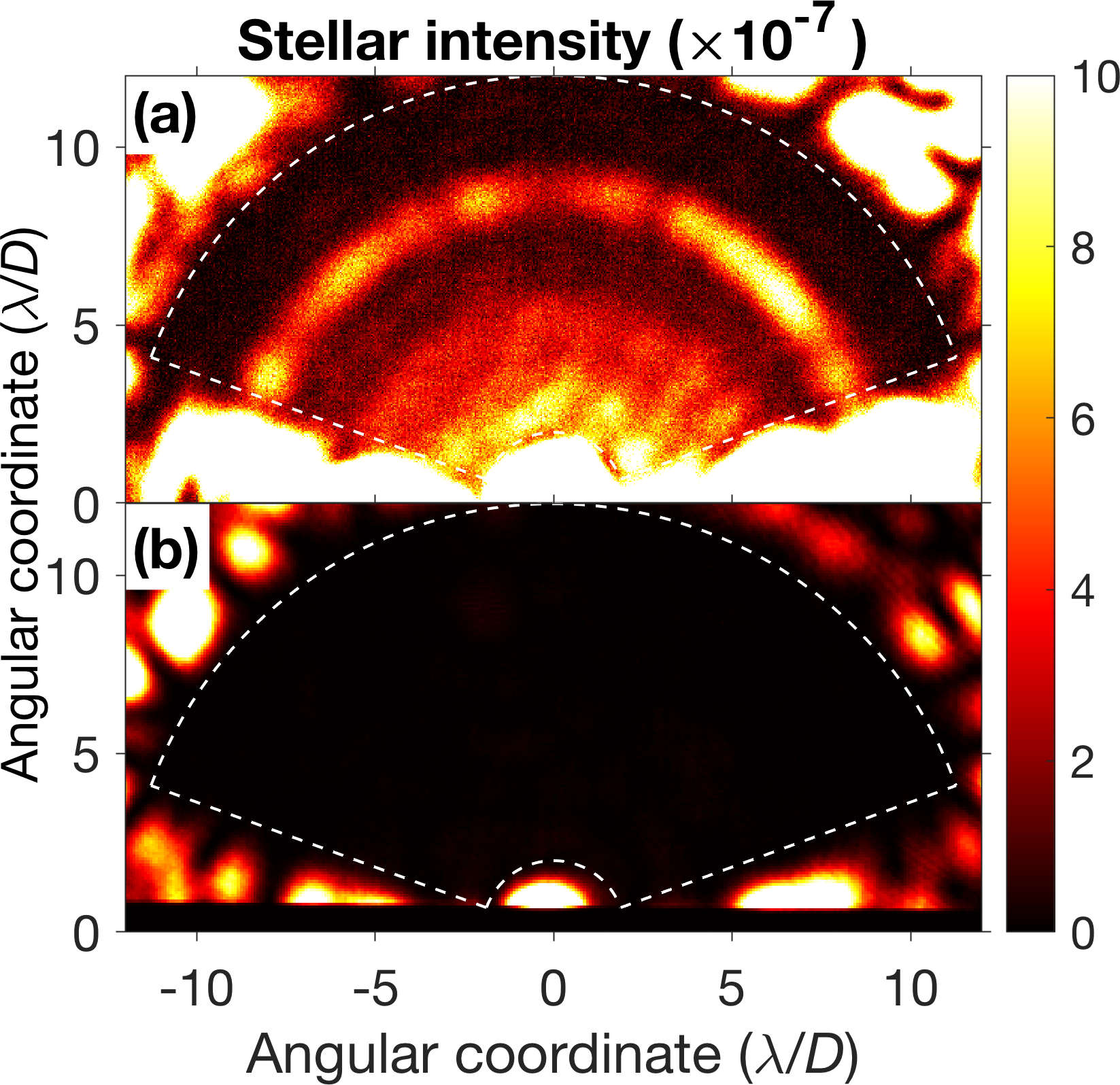}
    \caption{Normalized intensity in the experimental dark hole (a)~without and (b)~with the low pass filtered electronics. The pseudo-star is located at the origin in each case and the dashed lines shows the intended dark hole region, which is a partial annulus covering separations 2-12~$\lambda/D$ with a 140$^\circ$ opening.}
    \label{fig:DHs}
\end{figure}

Our first attempt at obtaining a dark hole with the new controller reached a contrast limit of $>$10$^{-7}$ in narrowband light ($\sim$20~nm bandwidth) centered at 600~nm. There was a distinct halo feature and a sharp ring of light at 8.7~$\lambda/D$ from the star with a contrast between 2$\times10^{-7}$ and $10^{-6}$ along the radius of the ring (see Fig.~\ref{fig:DHs}). We determined that the ring could not be modulated by the DM, which led us to the conclusion that it was incoherent with the light introduced by the DM probe shapes used for focal plane wavefront sensing. Through a number of tests, we ruled out external sources of light and stray light as the cause. We also arrived at a similar result using a longer coherence length Helium Neon (HeNe) laser source. 

\begin{figure}[t]
    \centering
    \includegraphics{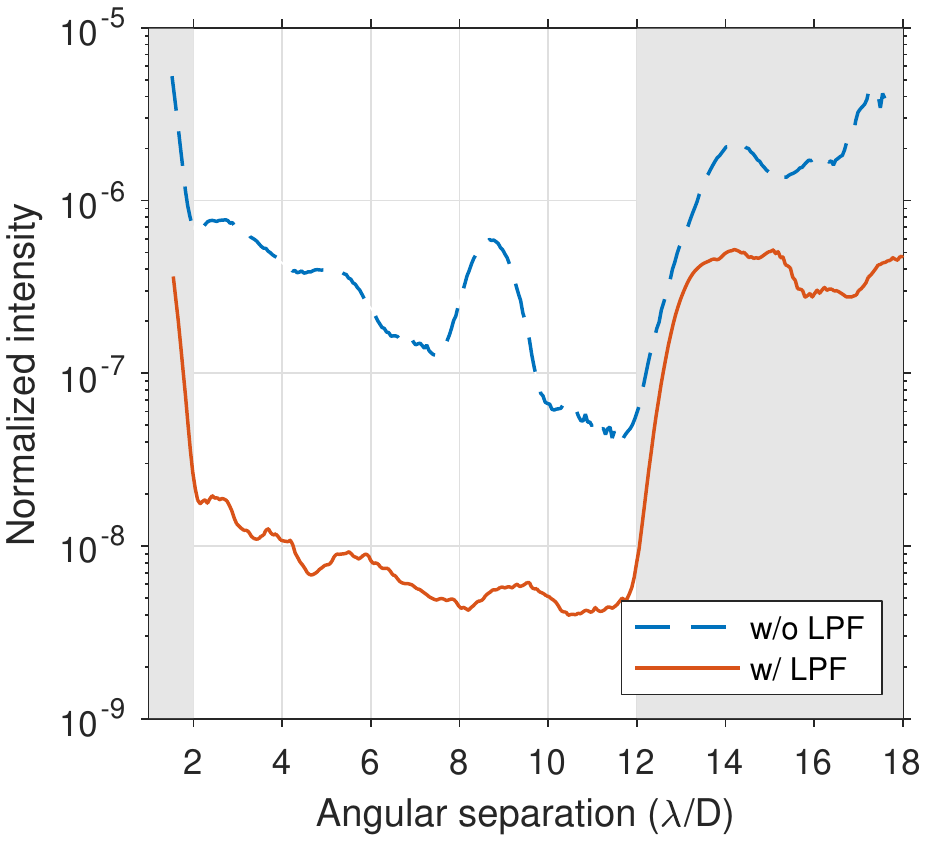}
    \caption{Azimuthal average of the normalized intensity within the dark hole regions in Fig.~\ref{fig:DHs} with and without the low pass filter (LPF) unit installed on the electronics.}
    \label{fig:radpros}
\end{figure}

The brightness of the incoherent signal did, however, respond to the air pressure in the chamber; specifically, increasing the air pressure to 100~Torr led to an order of magnitude improvement in contrast. We hypothesized that the air dampened fast vibrations in the DM surface membrane due to noise in the signal from the DM electronics, which created a temporally incoherent signal in the dark hole. To test this, we built a second version of the DM electronics, but this time we included low pass filters (LPF) with 20~M$\Omega$ resistance and 500~pF capacitance and tested them in a similar configuration. This resulted in a drastic improvement to the contrast, which we were able to reduce below 10$^{-8}$ over most of the dark hole region, with a mean of 5$\times$10$^{-9}$ over 2-12~$\lambda/D$ at similar wavelengths (see Fig.~\ref{fig:radpros}). Based on an empirical analysis on the impact of quantization errors, we estimated that this level of contrast requires a surface height resolution of 150 pm or smaller\cite{Ruane2020}. The RC filter has a time constant of 10ms, which dominates the controller response times for large amplitude commands. However, for operations in vacuum, that is when the filter is needed, commands are very small and response time is not critical, neither affected the experiment results. A future paper will detail our effort to mitigate the incoherent signal and provide further evidence to support our conclusions regarding the underlying mechanism.

\section{Picture-C implementation and first flight}\label{sec:picc} 

\begin{figure}[t]
    \centering
    \includegraphics[width=\linewidth]{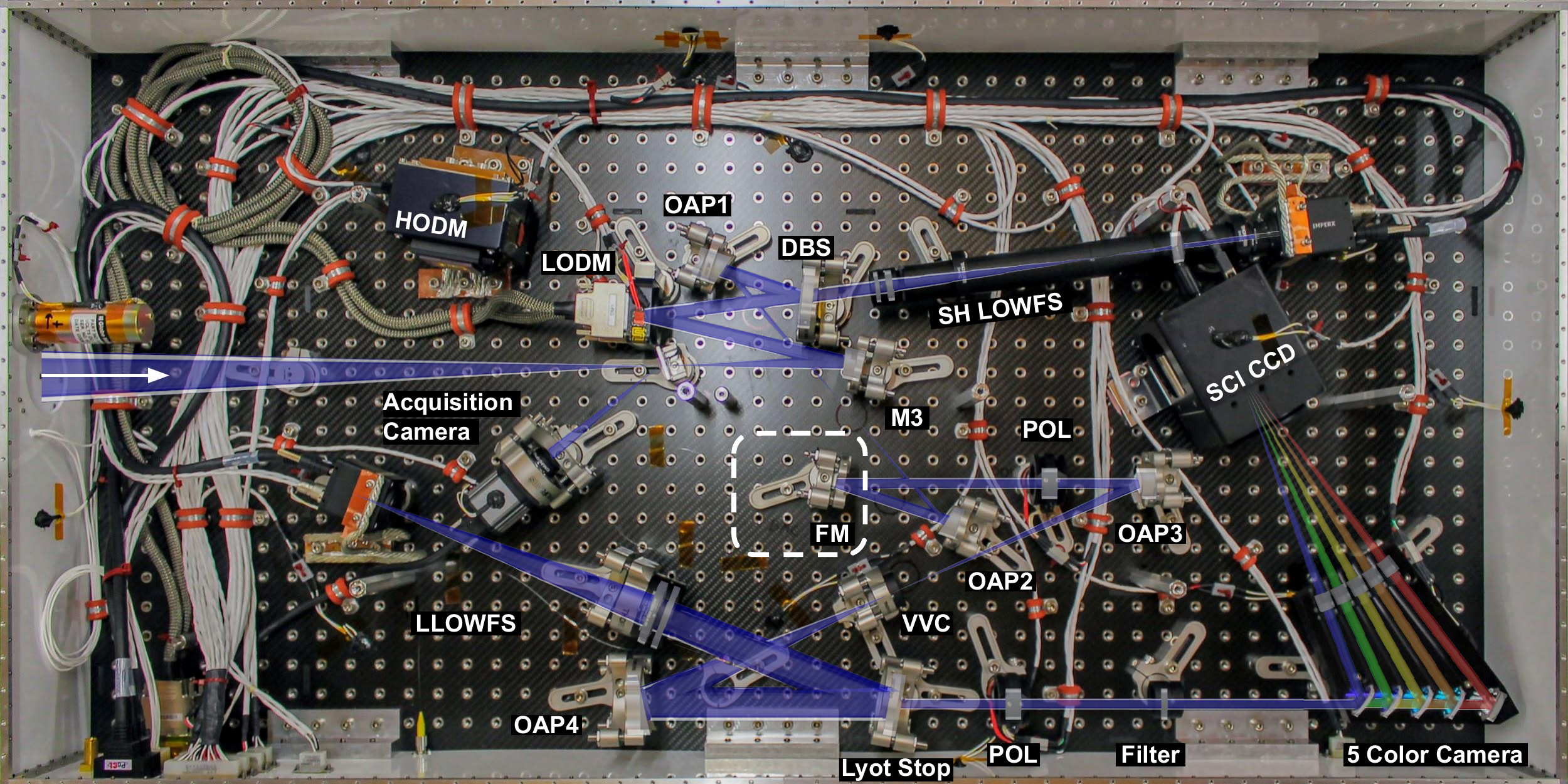}
    \caption{The PICTURE-C coronagraph is built on a 24"$\times$48" carbon fiber optical bench. Light from the telescope enters the instrument enclosure from the left. For the first flight, the high-order deformable mirror and controller (HODM) were relocated to the top left corner of the optical bench for testing outside the science beam path. The DM was replaced with a fold mirror (FM). The dashed square marks the intended DM location for the second flight.}
    \label{fig:piccoptics}
\end{figure}

\begin{figure}[t]
    \centering
    \includegraphics[trim={0 0 10cm 0}, clip,width=0.5\linewidth]{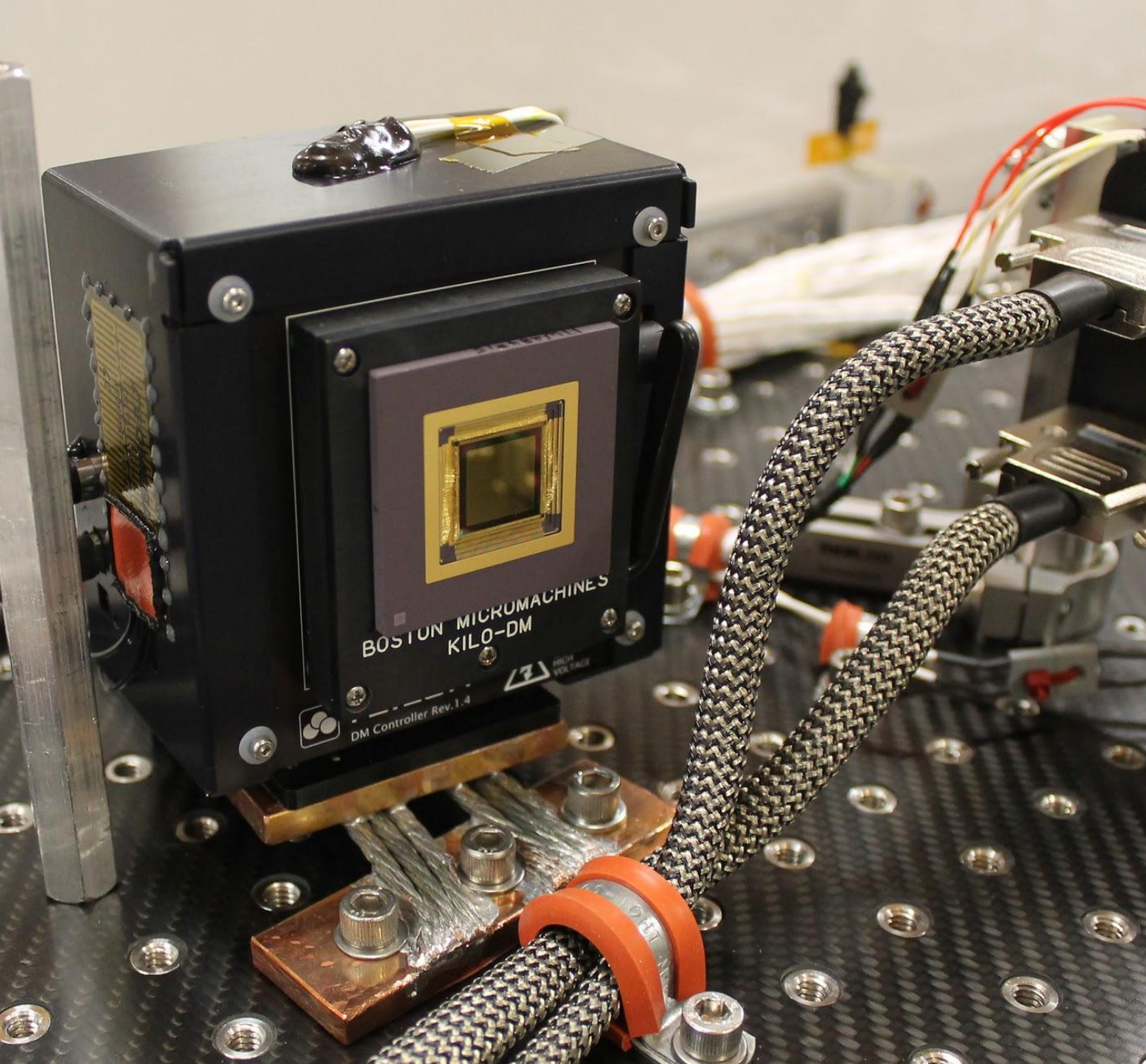}
    \caption{
    The PICTURE-C DM controller and BMC Kilo DM. Passive heat straps conduct heat into the optical bench. The controller was outfitted with two 10~W heaters to maintain the temperature at 20~C. 
    }
    \label{fig:piccdm}
\end{figure}

\subsection{Picture C overview}
The PICTURE-C high-altitude balloon mission\cite{Mendillo2019} aims to directly image debris disks and exozodiacal dust around nearby stars from a high-altitude balloon. The first flight of PICTURE-C launched from the NASA Columbia Scientific Balloon Facility in Ft. Sumner, NM, on September 28, 2019 and flew for a total of 20~hours, with 16~hours at float altitude above 110,000~ft. This flight successfully demonstrated key technologies for exoplanetary direct imaging missions and all hardware components for the second, science-focused, flight of PICTURE-C scheduled for the fall of 2020. These technologies include a vector vortex coronagraph, a high-speed low-order wavefront control system\cite{Mendillo2015,Hewawasam2017,Howe2017}, and the DM controller described here with a BMC Kilo DM. 

\subsection{DM Controller implementation}

The 952-actuator BMC Kilo DM and Mini controller will be used to implement high-order wavefront control for the second flight of PICTURE-C. For the first test flight, the DM and electronics were positioned outside of the beam path in an empty corner of the optical bench (see Fig.~\ref{fig:piccoptics}). The controller, Peltier cooler, and heatsink were removed for flight and replaced with passive heat straps that conducted heat into the optical bench (see Fig.~\ref{fig:piccdm}). During the flight, the controller ran a test program that held all actuators at 100~V bias and poked one random actuator at a time up to 150~V or down to 0~V at a rate of $\sim$1~Hz. The high-voltage bus was also cycled on and off $\sim$20 times.

\subsection{Flight results}
Pre- and post-flight tests were performed on the DMs and control electronics. The tests verified basic functionality of the controller, DM actuators, and controller channels. The post-flight test showed no changes to the DM or electronics; i.e. all DM actuators and controller channels remained functional after flight. The controller reported nominal housekeeping data (voltage, current, and temperature) for the duration of the flight at a rate of $\sim$1~Hz. Additionally, 11~high-voltage test points were commanded and monitored during flight and all reported correct voltages. The controller heatsink proved adequate for the flight environment and provided a slight cold bias during nighttime operation. Heaters provided 5-10~W (out of 20~W available) to keep the controller at 20$^\circ$C. During daytime operation the controller temperature rose only to 25$^\circ$C.

\section{Connectors and possible improvements}\label{sec:disc}

At the JPL HCIT facility, we operate and test multiple DM and coronagraphs. Actuator problems such as non-responsiveness, weak gains, and temporal variability have been traced several times to electronics and connectors rather than the DM itself. These problems have became a major issue during the testing and qualification of DMs for space, especially to assess if they can survive launch loads and vibration. The root cause of the problem is the number of connections, which scales with the number of actuators, and the additional connections needed for operation in vacuum. For example, a Kilo DM has roughly 1,000 actuators but for vacuum operation needs four connections: controller to ribbon cables, ribbon cables to vacuum feed-through, feedthrough to vacuum flex cables, and finally socket to DM. There are almost 4,000 connector pairs, so the probability of failure increases as the product of the number of actuators times the connections from the controller to the DM. If we repeat the same exercises with the 2K DM, we end up with 8,000 connector pairs. Since there is a need to increase DM acuator count for future missions, the logical path to improve reliability is to avoid connectors and miniaturize or solder every component. That is the approach taken with the Mini DM controller, which allowed us to have reliable, ``plug and play" units with 100\% actuator viability.

\begin{figure}[t]
    \centering
    \includegraphics[width=12cm]{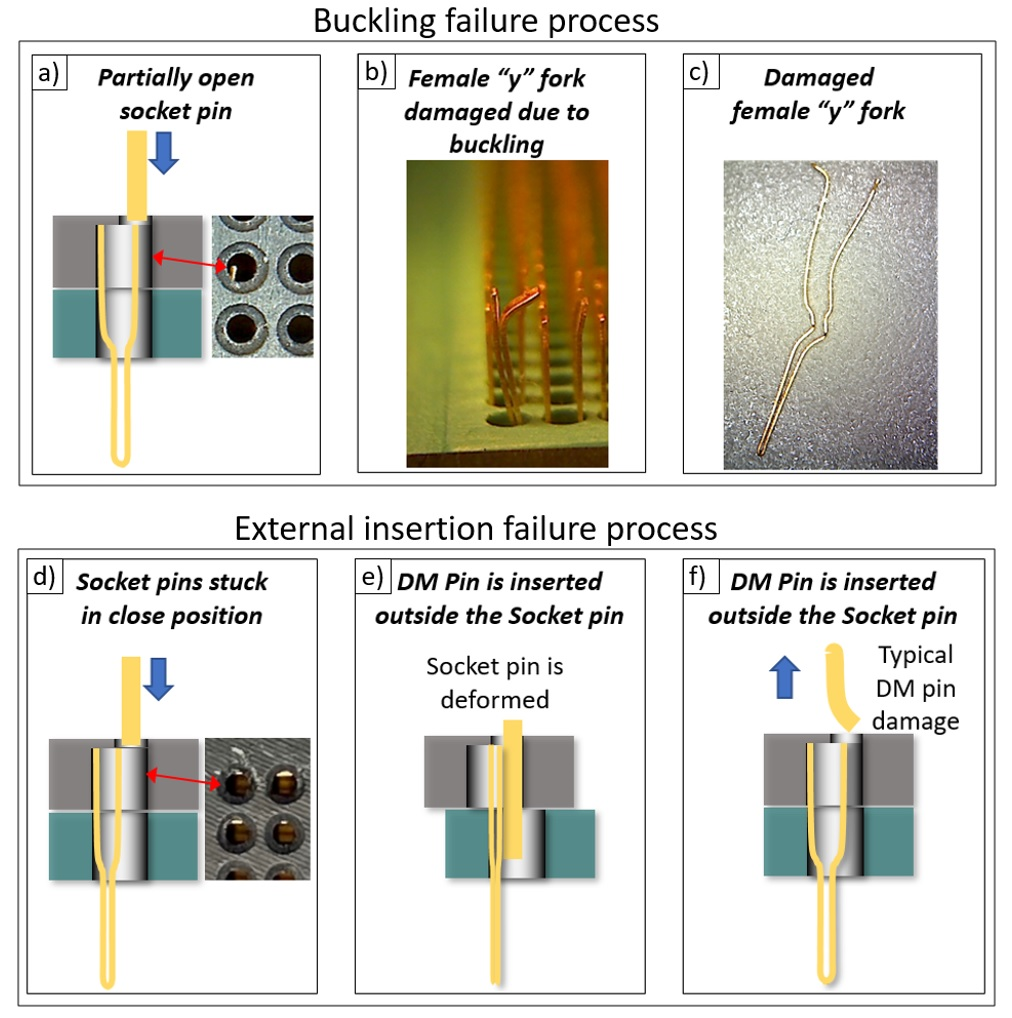}
    \caption{Socket insertion failure processes: (a)-(c)~buckling  and (d)-(f)~external insertion. (a) A “y” female pin does not open completely and partially blocks the opening. If the DM pin is pushed against the female pin it will buckle causing the damage shown on (b) and (c). (d)~The female pin remains closed when the ZIF socket is opened as shown on (d). The DM pin enters in the socket hole without difficulties as shown on (e), but when the ZIF socket is closed, both, female and male pins get deformed as shown on (f), likely causing a buckling failure on the next socket cycle.}  
    \label{fig:socket_damage}
\end{figure}

Even with the Mini DM controller, there remains a particularly delicate connector--the DM ZIF socket, which has been reported in the literature before \cite{Morzinski2012}. For the case of the Kilo DM, the manufacturer provides a socket made by Tactic Electronics. The socket uses a thin female ``y" shaped pin that clamps each DM channel pin when the lever to close the socket is actuated. We have observed two common failure modes for this socket including (1)~buckling, which occurs when the ``y" female fork does not open completely and the DM pin is inserted on top of it, and (2)~external insertion, which occurs when the ``y" fork does not open and the DM pin enters outside the fork. This results in a bent pin and fork after closing the socket. Buckling and external insertion problems are shown in Fig. \ref{fig:socket_damage}.

\section{Conclusions and Future work}\label{sec:conclusion}
Directly imaging exoplanets in reflected light is one of the next great frontiers in astronomy and has been identified in several of NASA's strategic documents. To observe Earth-like planets orbiting Solar-type stars requires contrasts on the order of 10$^{-10}$ at sub-arcsecond separations. Using an internal coronagraph with high-precision wavefront control to suppress unwanted starlight is a promising approach. The DM, a key component of a stellar coronagraph, is used to control the wavefront and suppress stellar speckles. 

Nowadays, MEMS DMs have a large actuator count, use high voltage, and have high accuracy and stability. These three factors, along with small production scales have resulted in only a few controller options that are bulky, massive, and not vacuum compatible. In addition, they require multiple connections between the unit and the DM, rapidly increasing the likelihood of a connection problem.

To solve this problem, we have developed a miniaturized DM controller that allows easy and reliable use of MEMS DMs, not only in air and vacuum test beds, but also in balloon and CubeSat missions. We established requirements that meet or exceed the performance of the most advanced devices available today, i.e.~16-bit resolution, adjustable dynamic range from 100V to 225V, resolution preservation, and 100\% actuator yield. In addition, we required the controller to be vacuum compatible, to operate in cold environments encountered on balloon missions, and to fit in a 1U CubeSat format with a 1~kg maximum weight and 8~W maximum power consumption.

The developed controller unit uses an array of 11 commercial HV-DACs soldered together with the DM socket. The real time electronics uses a rigid-flex PCB that allowed folding the board into 4 layers to meet the volume requirements. To enable vacuum operation, we applied underfill and conformal coatings to the electronics. We built a flight unit and spare for PICTURE-C. The flight unit went through thermal and vacuum testing at NASA Ames at the Engineering Evaluation Laboratory. Later, the unit was shipped to JPL, were it was tested for performance using a vortex coronagraph at the HCIT. We achieved a mean of 5$\times$10$^{-9}$ contrast over 2-12~$\lambda/D$ after including RC filters to dampen the DM membrane vibration. 
Later the unit flew onboard PICTURE-C for survival testing in preparation for the second operational flight. Telemetry showed nominal performance, and the unit returned to earth intact and with the same functionality. Recently, a third unit has been procured and is currently in use at the HCIT. 

In the future, we aim to develop a unit with a directly soldered DM to avoid the ZIF socket problems. In addition, we are exploring the possibility of producing a rad-hard version of the HV-DACs in order to build a flight unit capable of longer LEO missions or operating beyond LEO.

\acknowledgments     
The design and development of the first unit was funded by a NASA Ames Research Center Innovation Fund (CIF). The flight units were funded by NASA headquarters as PICTURE-C subaward to NASA Ames Research Center. The last unit was procured by JPL, and its performance characterization was carried out at the Jet Propulsion Laboratory, California Institute of Technology, under contract with the National Aeronautics and Space Administration (NASA).
\newline

\bibliography{refLibrary}   
\bibliographystyle{spiebib}   

\end{document}